\documentclass[twocolumn,citeautoscript,pra,superscriptaddress,amsmath,amssymb,8pt]{revtex4-2}

\usepackage{graphicx}
\usepackage{upgreek}
\usepackage{float}
\usepackage[T1]{fontenc}
\usepackage[colorinlistoftodos]{todonotes}
\usepackage{hyperref}
\hypersetup{
	colorlinks=true,
	linkcolor=blue,
	filecolor=blue,      
	urlcolor=blue,
	citecolor=blue
}

\newcommand{\abs}[1]{\ensuremath{\left| #1 \right|}}

\newcommand{\NN}{\text{uPINN }}

\begin{document}

\title{Untrained physically informed neural network for image reconstruction of magnetic field sources}

\author{A. E. E. Dubois}
\affiliation{Department of Physics, University of Basel, Klingelbergstrasse 82, Basel CH-4056, Switzerland.}

\author{D. A. Broadway}
\email{davidaaron.broadway@unibas.ch}
\affiliation{Department of Physics, University of Basel, Klingelbergstrasse 82, Basel CH-4056, Switzerland.}

\author{A. Stark}
\affiliation{QNAMI AG, Hofackerstrasse 40 B, Muttenz CH-4132, Switzerland}

\author{M. A. Tschudin}
\affiliation{Department of Physics, University of Basel, Klingelbergstrasse 82, Basel CH-4056, Switzerland.}

\author{A. J. Healey}
\affiliation{School of Physics, University of Melbourne, VIC 3010, Australia}
\affiliation{Centre for Quantum Computation and Communication Technology, School of Physics, University of Melbourne, VIC 3010, Australia}

\author{S. D. Huber}
\affiliation{Institute for Theoretical Physics, ETH Zurich, CH-8093, Switzerland}

\author{J.-P. Tetienne}
\affiliation{School of Science, RMIT University, Melbourne VIC 3000, Australia}

\author{E. Greplova}
\affiliation{Kavli Institute of Nanoscience, Delft University of Technology, 2600 GA Delft, The Netherlands}

\author{P. Maletinsky}
\email{patrick.maletinsky@unibas.ch}
\affiliation{Department of Physics, University of Basel, Klingelbergstrasse 82, Basel CH-4056, Switzerland.}

\begin{abstract}
	Predicting measurement outcomes from an underlying structure often follows directly from fundamental physical principles. However, a fundamental challenge is posed when trying to solve the inverse problem of inferring the underlying source-configuration based on measurement data. A key difficulty arises from the fact that such reconstructions often involve ill-posed transformations and that they are prone to numerical artefacts. Here, we develop a numerically efficient method to tackle this inverse problem for the reconstruction of magnetisation maps from measured magnetic stray field images. Our method is based on neural networks with physically inferred loss functions to efficiently eliminate common numerical artefacts. We report on a significant improvement in reconstruction over traditional methods and we show that our approach is robust to different magnetisation directions, both in- and out-of-plane, and to variations of the magnetic field measurement axis orientation. While we showcase the performance of our method using magnetometry with Nitrogen Vacancy centre spins in diamond, our neural-network-based approach to solving inverse problems is agnostic to the measurement technique and thus is applicable beyond the specific use-case demonstrated in this work.
\end{abstract}

\maketitle
\begin{figure*}[ht]
  \centering
    \includegraphics[width=0.8\textwidth]{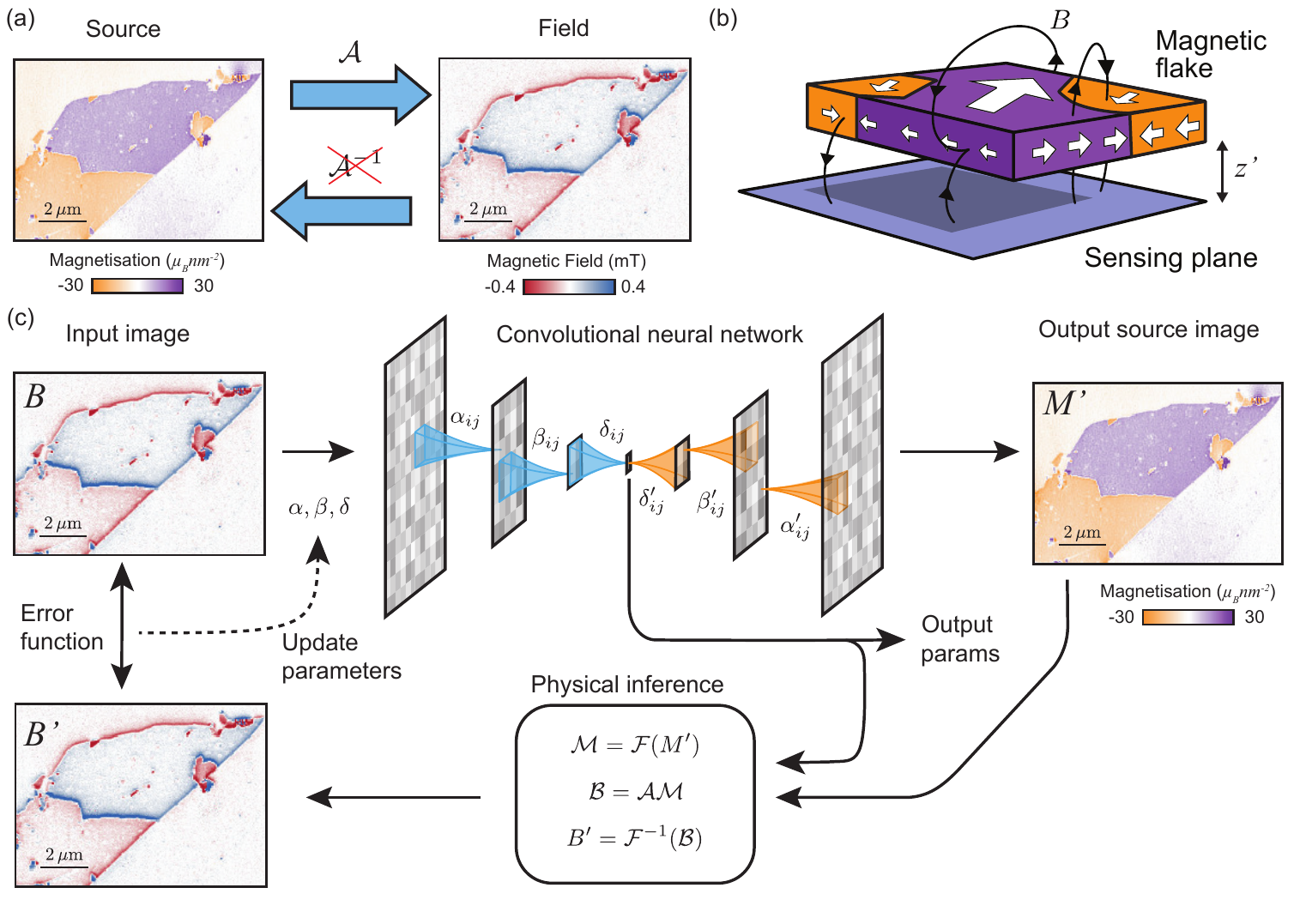}
    \caption{\textbf{Direct training method for solving inverse problems.}
    (a) Illustration of a source and field reconstruction problem, where the transformation $\mathcal{A}$ is well defined from source to field but the inverse, $\mathcal{A}^{-1}$ is not. 
    (b) Illustration of measurement of a magnetic field in a sensing plane offset from a magnetic material.
    (c) Diagram of untrained physically informed neural network (\NN). Where an input field image is put through a convolutional NN, with weightings at each layer, $\alpha, \beta, \delta$, etc. The reconstructed source image is transformed back into a field image with the well defined forward transformation, $\mathcal{A}$, and this reconstruction is used to generate the error function.}
    \label{Fig: introduction}
\end{figure*}

Determining the nanoscale magnetic state of materials is crucial to developing a deeper understanding of classical and quantum magnetism\,\cite{Casola2018} and the realisation of next generation  spintronics technologies\,\cite{Wiesendanger2016,Burch2018}. 
Typically, this information is difficult to achieve directly and requires, e.g., the use of large-scale imaging facilities\,\cite{Donnelly2017a}. 
An alternative method is to measure the fields emitted from the source, e.g. magnetic fields, and use these fields to infer information about the structure of the source. 
However, such a reconstruction process often involves solving an inverse problem that can be ill-posed and thus prone to significant errors (Fig.\,\ref{Fig: introduction}a). 
Despite this, by using appropriate assumptions and measurement configurations, the method of direct inversion has been used to probe different regimes of current flow\,\cite{Tetienne2017,Ku2020,Vool2020} and magnetisation in 2D van der Waals materials\,\cite{Broadway2020b,Thiel2019}.
However, extending these measurements to more exotic materials with complex magnetic structures becomes an issue as the error from the ill-posed transformation can become larger than the signal itself. 

In this work, we develop a new methodology for addressing this challenge.
We propose to facilitate the ill-posed transformation using neural networks (NN), which learn using a well-posed physically informed forward transformation.
We demonstrate that with our method, many of the issues that arise from traditional approaches can be  overcome.
While deep learning has already been employed to address inverse problems\,\cite{Lucas2018a,Prato2008,Ongie2020, Khoo2019,Arridge2019}, this approach has up until now relied on training of the network with large known data sets. 
This can result in excellent reconstructions, however, in fields that do not generate large swaths of data this type of technique is not applicable. 
While it is sometimes possible to simulate data for training, this inherently carries the risk of training the network to only solve a subset of the total number of problems and thus may converge to biased or non-physical solutions.
Likewise, training NNs on subsets of the total number of problems can result in non-physical solutions and a lack of generalisability.
Conversely, our method does not require prior training, as the forward transformation allows the network to learn the transformation for a given data set.
As such this is a broadly applicable technique for solving ill-posed reverse problems when the forward problem is well defined.

The relationship between 2D sources of magnetisation and magnetic fields can be described in the Fourier space by the following expression \cite{Broadway2020a}: 

\begin{align}\label{Eq 1}
    \begin{bmatrix}
    \mathcal{B}_x \\
    \mathcal{B}_y\\
    \mathcal{B}_z
    \end{bmatrix} 
    = \underbrace{\frac{-1}{\alpha}
    \begin{bmatrix}
    k^2_x/ k \quad  &k_xk_y/k \quad  &ik_x\\
    k_xk_y/k \quad &k^2_y/k \quad &ik_y \\
    ik_x \quad &ik_y \quad &-k
    \end{bmatrix}}_{\mathcal{A}}
    \begin{bmatrix}
    \mathcal{M}_x \\
    \mathcal{M}_y \\
    \mathcal{M}_z
    \end{bmatrix} 
\end{align}
where $\mathcal{B}$ and $\mathcal{M}$ are the Fourier-transformed magnetic field and magnetisation vector respectively, $\mathcal{A}$ is the transfer matrix, $k_x$, $k_y$ are the Fourier space coordinates with $k = \sqrt{k_x^2 + k_y^2}$, and $\alpha = 2 e^{kz'}/\mu_0$ contains an exponential that represents the propagator between the source and measurement planes \textcolor{blue}{and $\mu_0$ is the Vacuum permeability. }
Where we use the assumptions that the magnetisation vector, $\mathbf{M}(x,y)$, is confined to a 2D plane and that the stray field is measured in a parallel plane at an approximately known height $z'$, $\mathbf{B}(x,y,z')$ (Fig.\,\ref{Fig: introduction}b).

The transformation from a magnetisation map to a magnetic field map is a well-posed forward transformation with a unique solution.  
In contrast, the inverse problem of transforming from a magnetic field map to a magnetisation map is ill-posed as the transformation matrix  $\mathcal{A}$ is singular (det$(\mathcal{A}) = 0$), such that there is an infinite number of solutions for $\mathbf{M}$. The problem can be simplified using prior knowledge, e.g. one often assumes the magnetisation has a uniform known direction defined by spherical angles ($\theta,\phi$), i.e. $\mathbf{M}(x,y)= M_{\theta,\phi}(x,y) \mathbf{\Theta}$ where $\mathbf{\Theta} = (\sin \theta \cos \phi,\sin \theta \sin \phi,\cos \theta)$. The magnetic field measurement is also generally performed along a single direction $\mathbf{\Theta}_m = (\sin \theta_m \cos \phi_m,\sin \theta_m \sin \phi_m,\cos \theta_m)$, with measured projection $B_m$. Eq.~\ref{Eq 1} then simplifies to
\begin{align}\label{Eq 2}
	\mathcal{B}_m  = 
	\mathbf{\Theta_m} \cdot
	\mathcal{A}    
	\cdot \mathbf{\Theta}
	\mathcal{M}_{\theta,\phi} ,
\end{align}
where the transfer function is now $\mathcal{A}^\prime = \mathbf{\Theta_m} \cdot
\mathcal{A}    
\cdot \mathbf{\Theta}$ (which we will denote as simply $\mathcal{A}$ in the following).
Even so, this simplified problem is not invertible for certain values in k-space, e.g for $k = 0$, $k_{x,y} = 0$, and $k_x = -k_y\tan\phi $.
Moreover, some parameters entering the transfer function may be \textit{a priori} unknown, e.g. for an in-plane magnet ($\theta=90^\circ$ but $\phi$ unknown).

Traditionally, there exist two approaches to tackle this type of reconstruction. 
First, the analytical reconstruction tries to model the transformation $\mathcal{A}^{-1}$ using an optimization criterion and prior knowledge. 
For instance, regularization\,\cite{Meltzer2017} is an analytical model that attempts to solve the ill-posed problem by minimising an auxilary parameter that is added to remove the undefined transformation terms at the potential cost of spatial resolution. 
Second, one can treat unknown values as being stochastic, where it is then possible to estimate the full posterior $\textit{p}(x|y)$ with a Bayesian inference\,\cite{Clement2019}.
In contrast, our method, which leverages a machine learning architecture to generate the mapping from field to source, bypasses the need to rely on the ill-posed transformation.  


Deep neural networks are characterized as universal approximators having the potential to compute any nonlinear function\,\cite{Nielsen2015}. 
Thus, they are good candidates in tackling reconstruction problems\,\cite{Lucas2018a, Prato2008,Ongie2020, Khoo2019}. 
Given a data set $\boldsymbol{x_i}$, the reconstruction task is modelled by
\begin{equation}
\boldsymbol{y_i} = g_{\alpha}(\boldsymbol{x_i}) + \epsilon 
\end{equation}
where $g_{\alpha}(\boldsymbol{x_i})$ is the NN model with parameter-set $\alpha$, and $\epsilon$ is the noise in the measurement. 
The goal is to develop a model $g_{\alpha}$ that will learn to operate as $\mathcal{A}^{-1}$. 

While deep learning is the state of the art tool for solving many tasks in natural language processing or computer vision, serious improvements are needed to completely surpass traditional methods for reconstruction tasks\,\cite{Lucas2018a}.
The main issue is that the knowledge provided to the model to learn is constrained to the training data set. 
In our case, the number of parameters (e.g. magnetisation direction $\theta$, $\phi$, magnetometer standoff $z'$, and magnetometer direction $\theta_{m}$, $\phi_{m}$) influencing the reconstructed magnetisation is high. 
Thus, it requires a large data set to train the network with every possible combination of control parameters.
As such, there is a risk of training the network to only solve a subset of the total number of problems which may lead to the network converging to non-physical solutions.
Moreover, the network is likely to encounter measurements not seen in the training data as well as the presence of unseen noise. 
This is particularly problematic as even tiny perturbations in the input can produce artefacts in the reconstructed output\,\cite{Ronneberger2015}. 
Another challenge for deep learning is that the solution is based on a "black-box" which leaves out any explanations of the reconstruction process. 
The estimations on new samples are statistical inferences based on previous learning without warranty on the new output. 

In our method we circumvent these issues by having the neural network learn directly on each image, see Fig.\,\ref{Fig: introduction}c. 
While this learning procedure is similar to traditional training of a neural network, i.e. the weights of the network are updated according to the loss function, it differs in that the learning itself is never used for a different image. 
Nor does the network learn from multiple datasets like traditional training. 
Our method removes the reliance on large data sets with ground truths for loss functions. 
Instead it introduces a data specific loss function that is defined by performing the well-posed forward transformation on the ML output, to transform it back into the initially measured experimental quantity.
Thus, the error is defined as the difference between the original magnetic field and the reconstructed one.
By tailoring the loss function in this manner prior training of the network is no longer required.
Futhermore, the output of the ML code is restricted to being a physically relevant result as it must reproduce the measured field. 
We call this method the untrained physically informed neural network (uPINN) as opposed to the traditional pre-trained model. 

As this method directly learns on a single image, we do not face computational burden. Moreover, we can significantly alleviate over-training risk which usually refers to the inability to generalize well on unseen data due to an over-learning of the training data set's statistical distribution.
Hence, this method can be used with convolutional NN as well as fully connected NN. 
We opted for the former as it converges faster and minimises noise in the output image.
In this work, we present results obtained with a convolutional NN following a ``U-network architecture''\,\cite{Ronneberger2015}, details are given in Appendix~\ref{App: CNN arch}.
We also note that similar methods have been employed to reconstruct holograms using simulations fed from the output of CNNs~\cite{Hossein2020}.


\begin{figure}
    \centering
    \includegraphics{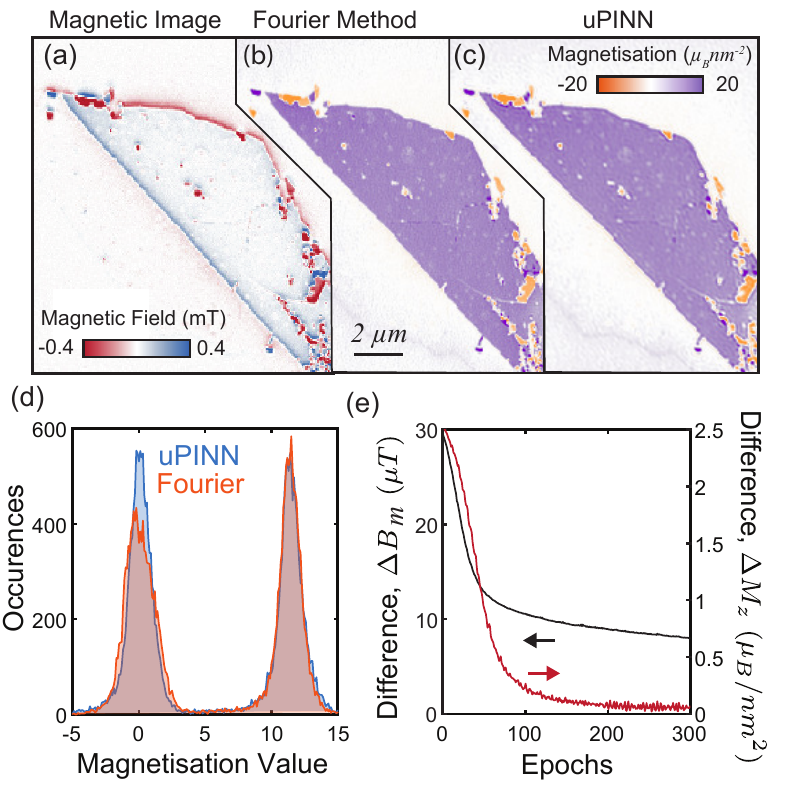}
    \caption{\textbf{Testing the neural network with out-of-plane magnetisation.}
    (a) Magnetic stray-field image ($\theta_m,\phi_m)$ $=(54.7^\circ, 282^\circ)$ of an out-of-plane magnetised tri-layer of the material CrI$_3$\,\cite{Thiel2019}. 
    (b) Magnetisation image produced using the traditional Fourier reconstruction method\,\cite{Broadway2020a}. 
    (c) Magnetisation image produced with the \NN. 
    (d) Histogram of the magnetisation values for the Fourier method (Orange) and the \NN (blue).
    (e) The mean difference between the original magnetic field and the \NN reconstruction as a function of number of epochs (left axis) and mean difference between the Fourier and \NN reconstruction methods (right axis).
    }
    \label{fig: converging}
\end{figure}

\begin{figure*}
	\centering
	\includegraphics{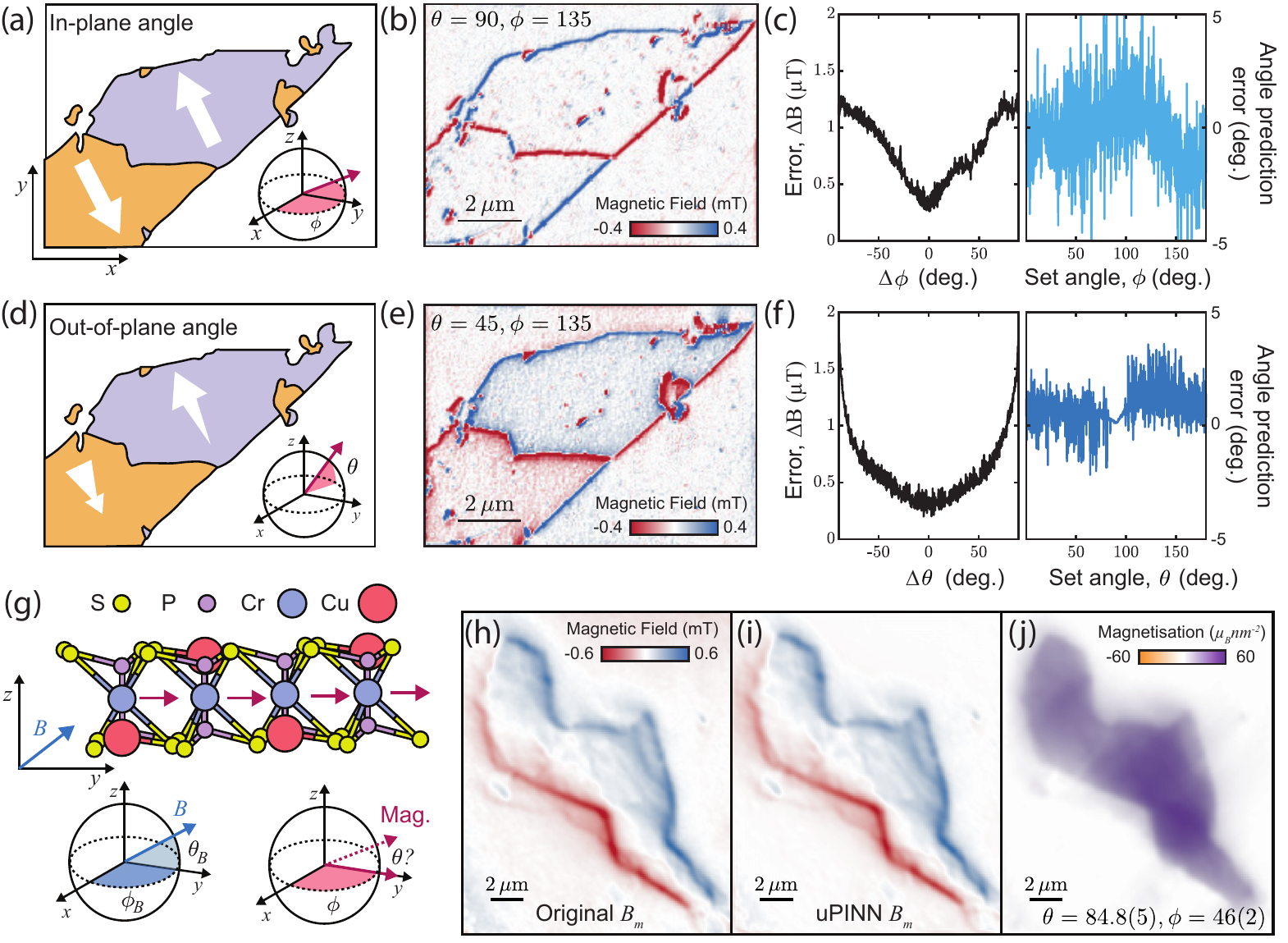}
	\caption{\textbf{Determining arbitrary magnetisation direction of easy-axis magnets with a neural network.}
		(a) Illustration of magnetisation image that is used to simulate magnetisation with an arbitrary direction. 
		(b) Example simulation of the magnetic field ($B_z$) image from a magnetisation direction of $(\theta, \phi) = (90^\circ, 135^\circ)$.
		(c) Comparison of the estimated in-plane angle ($\phi$) error from the \NN. Left panel: the error function evolution for different assumptions of angle showing a minimum for the correct assumption. Right panel: The error in the angle prediction for different actual angles.
		(d) Illustration of magnetisation canting out-of-plane to an arbitrary direction. 
		(e) Example simulation of the magnetic field ($B_z$) image from a magnetisation direction of $(\theta, \phi) = (45^\circ, 135^\circ)$.
		(f) Same as (c) but for out-of-plane ($\theta$) angles. 
		(g) Illustration of a monolayer of CuCrP$_2$S$_6$, which has a spontaneous magnetisation in-plane, with an applied magnetic field at  ($\theta_{B},\phi_{B}$) = ($54.7^\circ, 90^\circ$) canting the spins out of the plane with an unknown angle.
		(h) Magnetic image of a 300 nm thick CuCrP$_2$S$_6$ flake taken with a magnetometer and magnetic field direction of ($\theta_{m,B},\phi_{m,B}$) = ($54.7^\circ, 45^\circ$).
		(i,j) Reconstructed magnetic field (i) and magnetisation (j) of the CuCrP$_2$S$_6$ flake with the predicted magnetisation angle of $(\theta, \phi) = (84.8(5)^\circ, 46(2)^\circ)$.
	}
	\label{fig: Angle}
\end{figure*}

\begin{figure*}
	\centering
	\includegraphics{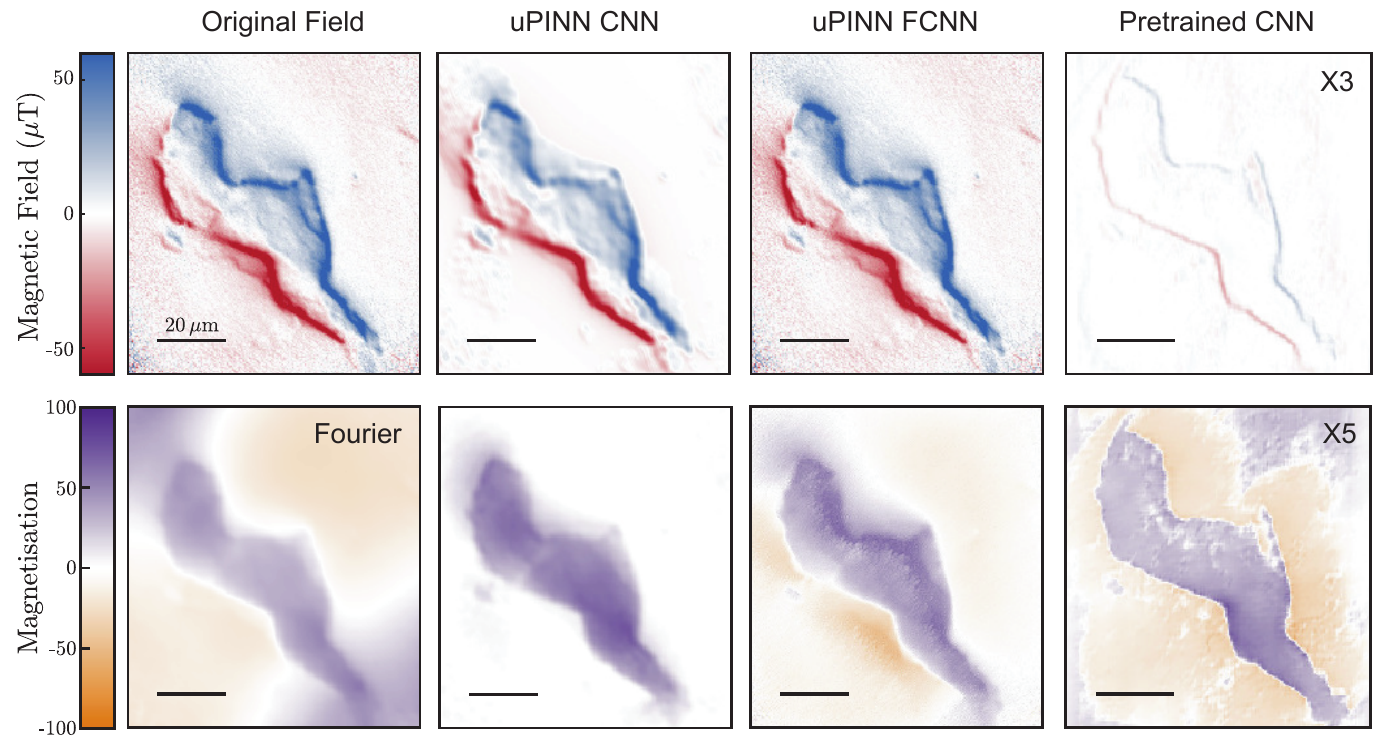}
	\caption{Comparison of different reconstruction methods on the same magnetic field image. The top row compares the original magnetic field to the reconstructed magnetic field from the neural network approaches :uPINN CNN, uPINN FCNN, trained CNN. The bottom row compares the magnetisation reconstruction from the traditional Fourier method and the same neural networks as the top row. The Fourier method include the use of spatial filter to minimise the noise~\cite{Broadway2020a} which may mask magnetisation information in more complex images.}\label{Fig comparison of models}
\end{figure*}

To benchmark our \NN with real data, we first employ it to a problem that is also manageable with regular methods, namely the reconstruction of an out-of-plane magnetised source i.e. $\theta = 0$ in Eq.~\ref{Eq 2}.
The data was taken using a scanning nitrogen vacancy (NV) magnetometer with an angle of ($\theta_m,\phi_m)$ $=(54.7^\circ, 282^\circ)$ \cite{Thiel2019} (Fig.\,\ref{fig: converging}a, b).
Our \NN quickly converged to a mapping that matches that of the standard method (Fig.\,\ref{fig: converging}c).
The two methods return the same average magnetisation of the material, $M_z = 11.5~\mu_B/$nm$^2$, while the \NN more accurately reconstructs the background magnetisation, observable in the height and width of the zero peak in the histogram in Fig.\,\ref{fig: converging}d.
We attribute this difference to a circumvention of some edge related artifacts that are generated in the Fourier method.

The Fourier method reconstructs based off the observed wave vectors in the magnetic image, as such it has a loss of information when the magnetic material is not fully contained in the image or if the magnetic field is truncated from the field of view. 
Our \NN method on the other hand does not have this restriction as it is fitting the image in order to reconstruct the magnetic field, as such truncation of the magnetic field does not directly affect the reconstruction. 
Likewise, source materials that cross the boundary of the image do not contribute to a significant loss in the quality of the reconstruction, as there is no magnetic field information to inform the fit to change from the observed source value. 
In principle, the \NN method could even be used to attempt to reconstruct sources that exists outside of the field of view but this aspect is beyond the scope of this work. 

We define a loss function, $\Delta B_{m} = \text{mean}(|B_{m} - B_{NN}|)$, that reliably estimates the quality of the reconstruction (Fig.\,\ref{fig: converging}e), reducing the average difference per pixel to $<8~$nT ($<3\%$ of the maximal magnetic field strength) within 300 epochs and is mainly localised to individual pixels rather than large areas.
Increasing the number of epochs eventually leads to the condition where the \NN begins to over fit the image and reconstructs the noise directly.
While the error function does not reflect this over training, it is visually evident and observed as an increase in the signal to noise ratio (SNR) of the output image (see Appendix~\ref{App: Noise} for details) and thus can be selected out \textit{a posteriori}. 
We compare the \NN reconstruction to that from the traditional method, $\Delta M_z = \text{mean}(|M_{\rm F} - M_{\rm \NN}|)$ (Fig.\,\ref{fig: converging}e, right axis), where $M_{\rm F}$ ($M_{\rm uPINN}$) is the reconstructed image via the fourier (uPINN) method. 
This comparison demonstrates that our model converges to a similar solution within 200-epochs, returning an average difference of $\Delta M_z \approx 0.07~\mu_B$nm$^{-2}$ ($<1\%$ of the maximal magnetisation strength) before becoming over-fitted, which is observed by the increase in the noise of the difference between the two methods.


We now move to solving the more difficult problem of reconstructing the magnetisation in a material with an arbitrary, but piecewise constant magnetisation direction.
These types of more general spin textures generate two issues in traditional reconstruction: First, the in-plane component of the magnetisation has more undefined terms than the case of purely out-of-plane magnetisation\,\cite{Broadway2020a}. 
Second, the introduction of the additional magnetisation angles greatly expands the parameter space. 
Here we take simulated data in order to reliably define the magnetisation angle (Fig.\,\ref{fig: Angle}a), and generate magnetic field images from which we try to reconstruct the underlying spin texture (Fig.\,\ref{fig: Angle}b).
To begin with, we confine the magnetisation to be purely in-plane $(\theta = 90^\circ)$, which is a valid restriction for materials with known easy-plane magnetic anisotropy. 
The \NN performs  well in reconstructing the simulated data with a high degree of accuracy and is capable of determining the in-plane angle. 
Performing this experiment 1000 times with random magnetisation angles (Fig.\,\ref{fig: Angle}c), demonstrates that the \NN does indeed accurately predict the angle, with an uncertainty of $<2\%$ for an SNR of 20.  

In some cases, the experimental conditions or the material quality may lead to a canting of the magnetisation direction (Fig.\,\ref{fig: Angle}d), resulting in the magnetisation having a component that is out of the plane (Fig.\,\ref{fig: Angle}e).
The \NN is able to determine the degree of this canting (Fig.\,\ref{fig: Angle}f) with an uncertainty of <1\% for an SNR of 20. 

To demonstrate this capability, we perform magnetisation reconstructions on experimental data for an in-plane magnetised material (CuCrP$_2$S$_6$, Fig.\,\ref{fig: Angle}g) that was measured with a non-planar magnetic field~\cite{Healey2022}, resulting in an unknown canting of the magnetisation, and a magnetic field image shown in Fig.\,\ref{fig: Angle}h. 
The \NN reconstruction (Fig.\,\ref{fig: Angle}i,j) converges to a magnetisation image that produces an average difference in the magnetic fields of $\Delta B \sim 2~\mu$T which corresponds to 3\% of the RMS magnetic signal strength. 
Here the \NN predicts a in-plane angle of $\phi = 46(2)^\circ$ and an out-of-plane magnetisation angle of $\theta = 84.8(5)^\circ$.
The former nearly matches the in-plane direction of the applied magnetic field $\phi_B = 45^\circ$ which sets the magnetisation direction within the easy-plane of the sample. 
While the value determined for the out-of-plane magnetisation angle $\theta$ is indicative of the out-of-plane component of the applied magnetic field ($\theta_B = 54.7^\circ$) canting the spins out of being purely in-plane.
The spin canting scales with the strength of the applied magnetic field (Appendix~\ref{SI: spin canting}), indicating that this is indeed a true spin canting rather than an artefact from the reconstruction process.


We performed comparisons of the magnetisation reconstruction for several different networks: the convolutional neural network (CNN), the fully connected neural network (FCNN), as well as a trained CNN. 
We did not train a FCNN because of computational burden and its poor generalization abilities. 
In order to train the CNN, we created a dataset with randomised magnetisation shapes and domains, which was used with the forward solution to obtain magnetic field images with the associated ground truth. 
The value and direction of the reconstructed magnetisation is influenced by the value of the magnetic field as well as 5 parameters: the magnetisation angles $\theta$ and $\phi$, as well as the angle of the sensor (here the NV orientation) $\theta_{NV}$ and 4) $\phi_{NV}$ and the sample to sensor standoff, $d_{NV}$.
All these parameters were randomly generated in order to create different combinations for the training. 
Additionally, we included random noise to the magnetic field image. 
The comparison of the different network's reconstructions is shown in the Fig.\,\ref{Fig comparison of models}. 

The magnetisation obtained with untrained CNN is more resistant to noise compared to the untrained FCNN due to the filtering effects of convolutional layers. 
Despite being sharper, the magnetisation obtained with a trained CNN shows artefacts due to the presence of unseen noise, namely the gradient in the input image and produces a drastically different overall magnitude, which can be attributed to the reconstruction of inverse direction magnetisation outside of the material.
However, the sharpness of the trained model does indicate that with a large and diverse training dataset, a combination of training and the physically inferred lose function may produce a more accurate result.

In contrast, the traditional Fourier reconstruction method generally amplifies noise when reconstructing in-plane magnetisation which requires spatial filtering to minimise~\cite{Broadway2020a} and as such is prone to artefacts in the background that can be larger than the reconstructed magnetisation. 
In principle this background can be removed with knowledge of the system~\cite{Broadway2020b}, but this still requires high quality data and is unusable for magnetisation images that are either noisy or don't capture a significant known zero magnetisation region to normalise the image. 
While uPINN offer a robust method to minimise the effect of these artefacts while still reconstruction a physically relevant representation of the magnetisation.
Additionally, uPINN doesn't add a significant computational burden, with image processing taking a few minutes for larger (512x512 pixels) images, which as a post processing technique is an acceptable time frame.


We have developed a method for solving ill-posed inverse problems with neural networks that use the well posed forward problem. 
This method is comparable to traditional Fourier reconstruction methods when the ill-posed component is minimal, but greatly outperforms when more significant ill-posed terms are introduced into the transformation or when some parameters are \textit{a priori} unknown. 
We have demonstrated that our technique is capable of reconstructing the magnetisation of in-plane magnetic material, which hitherto have been difficult to reconstruct. 
Additionally, we have shown that the \NN method is capable of predicting the magnetisation angle for an arbitrary direction opening the possibility of detailed studies of changes in magnetisation direction.
Furthermore, our method can also be extended to other transformation problems, such as reconstructing current density.
Finally, we emphasise that, while we demonstrated this technique using data obtained with nitrogen-vacancy centres in diamond, the technique is agnostic to sensor and only requires the magnetic field map (of any projection).  
As such, our method readily applies to other nanoscale magnetometry techniques such as, e.g., scanning nano-SQUID magnetometry\,\cite{Vasyukov2013a}.

\textbf{Author contributions}
The machine learning architecture was designed by A.E.E.D and D.A.B with assistance from P.M., A.S, E.G., and S.D.H.
M.A.T. took the measurements of the CrI$_3$ sample for in-plane simulations and testing of the machine learning code. 
A.J.H and J.-P.T. took measurement of the CuCrP$_2$S$_6$ sample for testing the angle determination. 
A.E.E.D, D.A.B, and P.M wrote the manuscript.
All authors discussed the data and commented on the manuscript.


\bibliographystyle{apsrev4-2}
\bibliography{bib}


\appendix

\section{Regression task using a feed-forward fully-connected network}
A deep learning architecture is made of neurons that are stacked by layers. 
Where a neuron $z$ is a linear combination of a weight, $w$ and a bias $b$, such that $z = {x}_i{w}_1+{b}_1$ for a given input ${x}_i$.

A network that consists of a single hidden layer $l$ with $q$ neurons holds the following relationship
\begin{equation}
	{y}_i \approx g_\alpha(\boldsymbol{x}_i) = 
	f({x}_i{W}_1+{b}_1){W}_2
	+{b}_2
\end{equation}
where ${W}_1$ and ${W}_2$ are weight matrices of size $(d \times q)$ and $(q \times k)$, and the bias vectors ${b}_1$ and ${b}_2$ are of size $q$ and $k$ respectively. Where $d$ is the size of the input data set, $q$ is the number of neurons in the layer, and $k$ is the number of neurons in the output layer denoted with the subscript 2. 
The weights and bias vectors are the set of parameters $\alpha$ that the model has to update while training. 
The activation function $f(.)$ is a non linear transformation carried out on each neuron at each layer. A detailed description of this type of neural network can be found in Ref~\cite{Nabian2020}.

In general, the approximation of complex non-linear functions can be dramatically improved by increasing the number of neurons $z$ in each layer or by adding more layers $l$ \cite{Nielsen2015}.
In multiple layer architectures, the output of each activation function becomes the next layer input, which is then transformed by a new weight matrix and bias vector. 
At the end of such a multi-layer network, a specific activation function $f(.)$ - called output function- is used to define the output format into the desired form.
In the method presented in this work the output is an image that is the same size as the input image but this can also be used to restrict the output to possibilities for classification problems. 

\section{Training a neural network}
In order to train a network one needs a dataset made of pairs of inputs and outputs $({X},{Y})$. 
At the end of a forward pass, the network estimates an output $Y^\prime$ from the given input, $X$. 
The difference between the ground truth $Y^T$ and the estimation $Y^\prime$ is computed to obtain the loss function. 
A common loss function is the mean absolute error, given by
\begin{equation}
	C({Y^{T}},{Y^{\prime}}) = \frac{1}{n} \sum^n_{i=1}{ \abs{{g(x)}_i-{y}_i^{t}}},
\end{equation}
where $n$ is the number of output parameters, $g(x)_i$ is the network estimation for the $i$-th parameter, and $y_i^t$ is the corresponding ground truth. 
In order to update the network parameters, a backpropagation algorithm evaluates the gradients of the loss function $\nabla_\alpha C$ with respect to the weights and biases. 
This procedure begins calculating the gradients from the last layer and continues layer by layer using the chain rule, which qualifies how much the value of each parameters affects the final loss of the network. 

Once the gradients have been evaluated, the optimizer defines parameter update. Commonly, the simplest version of this optimizer can be used, the standard stochastic gradient descent. 
At each iteration $i$, it updates individual values of parameters $\alpha$ based on their respective gradients and a learning rate $\eta$.

\begin{equation}
	{\alpha^{i+1}} = \alpha^{i} - \eta^i  \nabla_\alpha C^i
\end{equation}

The whole process is set as an optimization problem 
$\arg\min_{\alpha} C$ where forward-feeding is repeated until some stopping criteria are met.

\section{Training and learning of a purely physically inferred neural network}

Usually, regularization acts on the weights $\alpha$ size in order to overcome over-fitting. 
Where a physics regularization deals with the implementation of knowledge into the model. 
This added knowledge constraints the learning to a desired physical solution by adding a penalty term $\Phi({Y^o})$ to the loss function $C$. 
In our case, this term is the forward solution for reconstructing the magnetic field from the magnetization.
Thus, at the end of each iteration, the magnetic field is computed from the current estimated magnetisation. 
By doing so, it can be compared with the original magnetic field $X^0$ at each iteration and be included in the loss function. 
The complete loss function with physics constraint $C_r$ takes the form of
\begin{equation}
	C_r({Y^t},{Y^o}) = C({Y^t},{Y^\prime}) + \lambda[C_p(X^0,\Phi({Y^\prime}))],
\end{equation}
where $\lambda $ is a hyper-parameter controlling the influence of the physics term and $C_p$ refers to the physic inference cost function. 
By included this addition term, one can ensure that even on a subset of all possible data, the model will learn the proper function. 

In the situation where no ground truths are available it is possible to get rid of the first part of the loss function all together, keeping only the physics term. 
The loss function then becomes
\begin{equation}
	C_r({Y^t},{Y^\prime}) = C_p(X^0,\Phi({g(X^0)}))
\end{equation}
The ground truth $Y^t$ disappeared from the loss function and this one only depends on the original input $X^0$ and the learned function $g(.)$ that gives the network estimation. 
This modification allows a model to train without the required ground-truth value. 
Moreover, it can directly learn on a single new input without previous training. 
So instead of traditionally training on a dataset and using the learned weights to make new predictions, a network updates its parameters ${\alpha}$ directly on the measured quantity $X^0$. 

The output of the network is not a statistical inference anymore but a solution depending of the well-posed forward transformation. 
Thus, the model's knowledge is not limited to a previous dataset but is directly taken from the input. 
Seeing that the loss function depends of the well-posed forward transformation, the proposed solution is ensured to be physically relevant.

\section{Model and network architecture}\label{App: CNN arch}

\begin{table*}
	\centering
	\begin{tabular}{c | c | c | c | c | c }
		& \, Layer Type \,  & \, Filters \, & \, Kernel Size \, & \, Stride \, & \, Neurons \, \\
		\hline
		&conv & 8 & 5 & 1 & na\\
		&conv (ROI)  & 8 & 5 & 1 & na\\
		&conv & 8 & 5 & 2 & na\\
		&conv & 16 & 5 & 2 & na \\
		&conv & 32 & 5 & 2 & na\\
		& conv & 64 & 5 & 2 & na \\
		Image&conv  & 128 & 5 & 2 & na\\
		&conv  & 64 & 5 & 2 & na\\
		&conv & 32 & 5 & 2 & na \\
		&conv   & 16  & 5 & 2 & na\\
		&conv & 8 & 5 & 2 & na\\
		&conv & 1 & 5 & 2 & na\\
		&conv & 1 & 3 & 1 & na\\
		\hline
		& Dense & na & na & na & 128\\
		Parameters & Dense & na & na & na & 64\\
		& Dense & na & na & na & 1\\
		\hline
	\end{tabular}
	\caption{Architecture of the convolutional neural network}
	\label{table: CNN architecture}
\end{table*}
\vspace{0.8\baselineskip}

The possibility to train the network directly on a single image makes it possible to use different architecture without facing a computational problem. 
In this main text, we used a convolutional neural network (CNN) following a U-net architecture. 
Where the cost function was the mean average loss between the reconstructed magnetic field and the original one. 
Seeing that convolutional layer requires a fixed-size input, a border is added to the input image to match the correct format. 
In order to make sure this border is not taken into account into the reconstruction, a region of interest (ROI) layer is inserted to focus the learning of the network to the desired region~\cite{Eppel2017}. 
Out of the 6 parameters influencing the reconstruction of the magnetisation, 4 can be deduced by the model directly. 
It consists of the 2 NV and magnetisation angles. For this purpose, a sub-network -fully connected- is forked from the main network for each one of those parameters. 
The architecture for the convolutional neural network is displayed in the Table~\ref{table: CNN architecture}. We also implemented a fully connected network (FCNN) to compare with, whose architecture is displayed in Table \ref{table: FCN architecture}.

\begin{table*}
	\centering
	\begin{tabular}{c | c | c | c | c | c }
		& \, Layer Type \,  & \, Filters \, & \, Kernel Size \, & \, Stride \, & \, Neurons \, \\
		\hline
		&Dense & na & na & na & 256\\
		&Dense & na & na & na & 128\\
		&Dense & na & na & na & 64\\
		&Dense & na & na & na & 32 \\
		Image & Dense & na & na & na & 16 \\
		&Dense & na & na & na & 32 \\
		&Dense & na & na & na & 64 \\
		&Dense & na & na & na & 128 \\
		&Dense & na & na & na & 256 \\
		\hline
		&Dense & na & na & na & 128\\
		Parameters &Dense & na & na & na & 64\\
		&Dense & na & na & na & 1\\
		\hline
	\end{tabular}
	\caption{Architecture of the fully connected neural network}
	\label{table: FCN architecture}
\end{table*}
\vspace{0.8\baselineskip}

\section{Noise regulation}\label{App: Noise}

One issue with implementing any reconstruction method is noise in the original data can be amplified during the reconstruction process. 
Thus data with excessive noise is often ruled as ineligible for reconstruction, requiring either additional measurement time or a new measurement to be taken.
This reconstruction is no different, in fact, the ill-posed problem is known to amplify noise~\cite{Broadway2020a}.
With a trained neural network, one can generalize a model to get rid of the noise in the reconstruction. 
However, you still face the risk that unseen perturbations creating artefacts in the reconstruction (see Fig.~\ref{Fig comparison of models}). 
The untrained neural network reconstructs the image without amplification of the noise, instead the noise is reflected in the reconstructed image directly. There is a trade-off here between the generalization power offered by a trained network and the accuracy and robustness offered by our direct network.

As the untrained neural network learns on the input directly, the quality of the reconstruction is directly linked to the quality of the input. 
Thus, we can easily improve the reconstruction quality by minimising the noise beforehand. 
However, removing noise always carries the risk of removing valuable information, which is particularly damaging when dealing with precise measurements. 
Isolating the denoising from the reconstruction is valuable, as it allows for the possibility of using noisy images without the need for additional training and facilitates double checking of the denoise process itself for loss of information. 
In contrast, a trained neural network model would have to learn the denoising process, and can thus generalise incorrectly for a given data set.

\begin{figure}
	\includegraphics[width=0.49\textwidth]{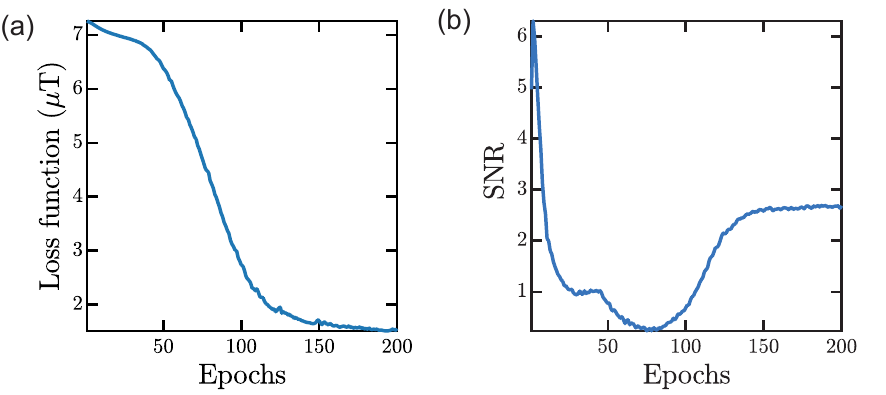}
	\caption{Error and SNR evolution during the reconstruction process}
	\label{SI Fig: error and SNR}
\end{figure}

In the comparison of models we see that CNN has the power to be more noise resistant compared to FCNN due to the convolutional layers. 
However, the longer the untrained model learns, the more noise is taken into account into the reconstruction. 
In order to avoid this situation, we include an early stopping criteria, a signal to noise ratio (SNR) metric. 
Once the SNR has reached a minimum, the learning stops even though the loss function can still be reduced. 
See Figure~\ref{SI Fig: error and SNR} for a typical behaviour of the loss function and SNR for increasing epochs.

\section{Determination of out of plane spin canting}\label{SI: spin canting}

\begin{figure*}[h]
	\includegraphics[width=0.95\textwidth]{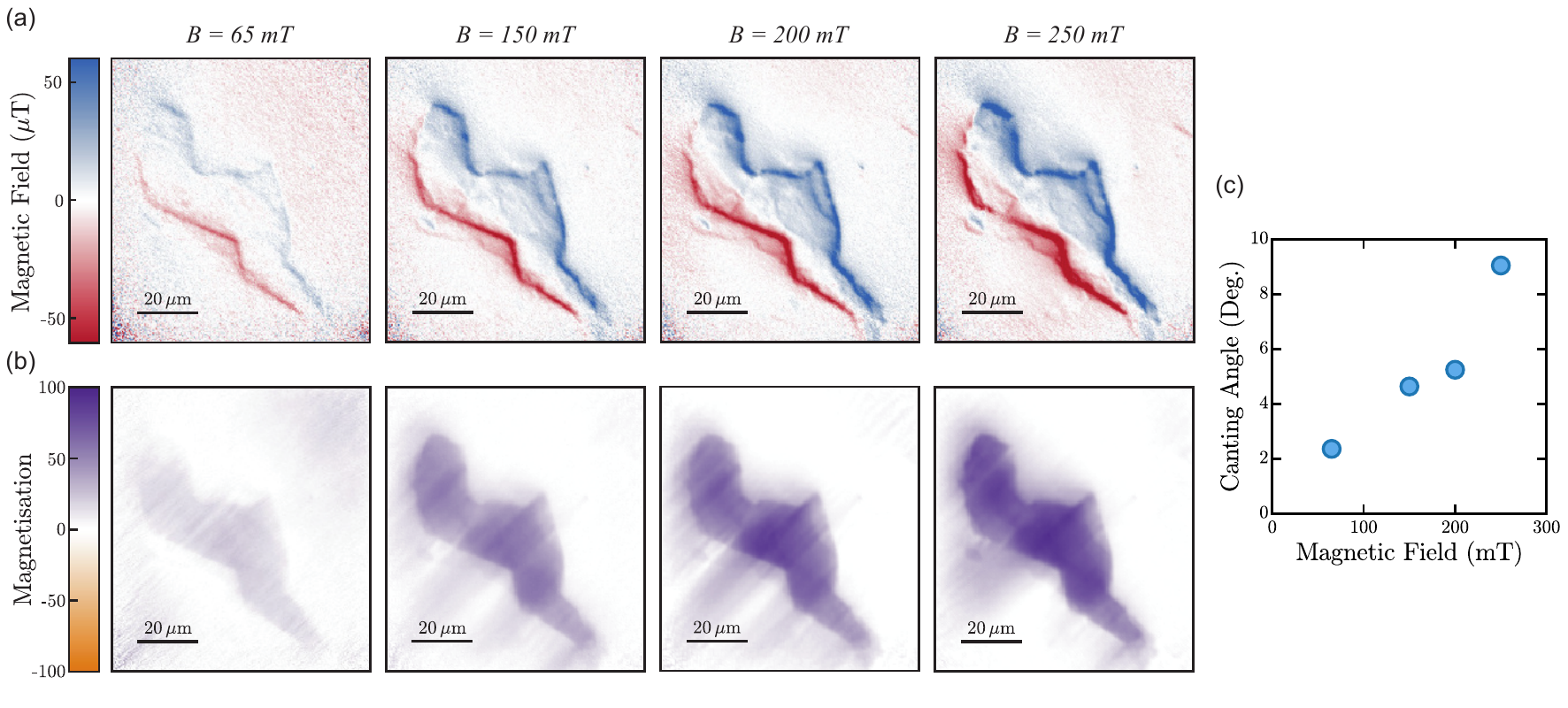}
	\caption{Determining out-of-plane spin canting with neural networks training.
		(a) Magnetic images of the same flake in the main text taken at different applied magnetic fields. $B = (65, 150, 200, 250)$~mT, where the field was applied along an in-plane angle of $B_\phi = 45^\circ$ and an angle from the z-axis of $B_\theta = 54.7^\circ$
		(b) Corresponding magnetisation reconstructions from the above magnetic fields. 
		(c) The predicted spin canting angle ($90^\circ - \theta$) from purely in-plane versus the applied magnetic field.
	}
	\label{SIFig: out-of-plane}
\end{figure*}

While the determination of the spin angle in simulations was relatively robust to noise, it is important to qualify how reliable the technique is on real data. 
Unfortunately, with datasets like these we don't have the ground truth. 
However, we can measure trends in the spin orientation to determine if these make physical sense.
Here we examine the reconstructed out-of-plane angle of CrCuP$_2$S$_6$ as a function of the applied magnetic field.

CrCuP$_2$S$_6$ is an in-plane A-type antiferromagnet, which in bulk has a small difference in susceptibility between the in-plane and out-of-plane direction~\cite{Colombet1982}. 
While DFT calculations show that in the monolayer limit there exists a weak anisotropy~\cite{Qi2018}.  
A naive single-domain shape anisotropy calculation (agreeing with monolayer DFT calculations) predicts negligible canting over the range of fields probed ($<2^\circ$ at $250$~mT) however it is unlikely to capture the complex interplay between the various anisotropies and exchange terms in thicker flakes. 
Our NN analysis shows a monotonic increase in the canting angle (see Fig~\ref{SIFig: out-of-plane}) as expected, but the increase is much more rapid than the naive prediction. 
As this is an A-type antiferromagnet and is relatively thick (
$\sim 300$ layers) it is possible that the monolayer DFT calculations are no longer valid in this regime. 

Precise measurement of magnetisation angle is difficult to obtain by other means thus this demonstrates the power of this NN approach to determine small perturbations of spin alignments.

\end{document}